\documentstyle[prb,aps,twocolumn,epsf,floats]{revtex}
\def\geqap{\,\raise 2pt \hbox{$>\kern-11pt \lower 5pt \hbox{$\sim$}$}\,}
\def\leqap{\,\raise 2pt \hbox{$<\kern-10pt \lower 5pt \hbox{$\sim$}$}\,}
\begin{document}
\draft
\twocolumn[\hsize\textwidth\columnwidth\hsize\csname @twocolumnfalse\endcsname
\title{Orbital Degree of Freedom and Phase Separation \\ 
in Ferromagnetic Manganites at Finite Temperatures}
\author{S. Okamoto, S. Ishihara, and S. Maekawa}
\address{Institute for Materials Research, Tohoku University,  Sendai,
980-8577 Japan}
\date{\today}
\maketitle
\begin{abstract} 
The spin and orbital phase diagram for perovskite manganites are investigated 
as a function of temperature and hole concentration. 
The superexchange 
and double exchange interactions dominate the ferromagnetic 
phases in the low and high concentration regions of doped holes, respectively. 
The two interactions favor different orbital states each other. 
Between the phases,  two interactions compete with each other
and the phase separation appears in the wide range of temperature and 
hole concentration. 
The anisotropy of the orbital space causes discontinuous 
changes of the orbital state and  
promotes the phase separation.  
The relation between the phase separation and the 
stripe- and sheet-type charge segregation
is discussed. 
\end{abstract}
]
\narrowtext
\section{Introduction}
Doped perovskite manganites and their related compounds 
have attracted much attention, 
since they show not only the colossal 
magnetoresistance (CMR) \cite{chahara,helmolt,tokura,jin}
but many interesting phenomena 
such as a wide variety of magnetic structure, 
charge ordering and 
structural phase transition. 
Although the ferromagnetic phase 
commonly appears in the manganites, 
the origin still remains to be clarified. 
Almost a half-century ago,  
the double exchange (DE) interaction 
was proposed to explain the close connection 
between the appearance of ferromagnetism 
and the metallic conductivity. \cite{zener,anderson}
In the scenario, 
the Hund coupling between carriers and localized 
spins is stressed. 
It has been recognized that the ferromagnetic metallic state 
in the highly doped region 
of La$_{1-x}$$A_x$MnO$_3$ ($x \sim 0.3$) with $A$ being a divalent ion 
is understood based on this scenario, 
where the compounds show the wide band width. \cite{kubo,furukawa}
\par
On the contrary, 
the DE scenario is not applied to 
the lightly doped region \cite{millis}
$(x<0.2)$ where the CMR effect is observed. 
In the region, the degeneracy of $e_g$ orbitals in a 
$\rm Mn^{3+}$ ion exists and affects the physical 
properties. 
The degeneracy is called the orbital degree of freedom. 
With taking into account the orbital degree together with 
the electron correlation, 
the additional ferromagnetic interaction, 
that is, the ferromagnetic superexchange (SE) 
interaction, is derived. 
This is associated with the alternate alignment of the 
orbital termed antiferro(AF)-type 
orbital ordering. \cite{goodenough,kanamori,kugel}
The SE interaction dominates the ferromagnetic spin 
alignment observed in the $ab$-plane in $\rm LaMnO_3$
and the quasi two-dimensional dispersion relation of the spin wave 
in it. \cite{hirota,ishihara1}
When holes are introduced into the insulating $\rm LaMnO_3$, 
the successive transitions occur in magnetic and transport phase diagrams; 
with increasing $x$, it is observed in La$_{1-x}$Sr$_x$MnO$_3$ as 
almost two dimensional ferromagnetic (A-type AF) insulator  
$\rightarrow$ 
isotropic ferromagnetic insulator 
$\rightarrow$
ferromagnetic metal.\cite{tokura,kawano}
The first order phase transition 
between two ferromagnetic states 
recently discovered in La$_{1-x}$Sr$_x$MnO$_3$ with 
$x \sim 0.12$ \cite{endoh} 
indicates that the orbital state also changes 
at the transition. 
In order to understand a dramatic change of electronic states 
in lightly doped region and its relation to CMR, 
it is indispensable to study the mutual relation 
between the two ferromagnetic interactions, i.e., DE and SE. 
\par
In this paper, we investigate the 
spin and orbital phase diagram 
as a function of temperature $(T)$ and hole concentration $(x)$. 
We focus on the competition and cooperation between 
the two ferromagnetic interactions SE and DE.  
We show that the SE and DE interactions dominate the ferromagnetic phases 
in the low and high concentration regions of doped holes, respectively, 
and favor the different orbital structures each other. 
Between the two phases, 
the phase separation (PS) appears in the wide range of $x$ and $T$. 
It is shown that the phase separation is promoted by 
the anisotropy in the orbital space.
The spin and orbital phase diagram at $T=0$ 
was obtained by the Hartree-Fock 
theory and interpreted in terms of the SE and DE interactions 
in Ref. \onlinecite{maezono}.  
The PS state between 
two ferromagnetic phases driven by the DE 
interaction and the Jahn-Teller distortion 
at $T=0$ was discussed in Ref. \onlinecite{yunoki2}. 
In this paper, we obtain the PS state based on the model with 
strong correlation of electrons at finite $T$. 
\par
In Sect.~II, 
the model Hamiltonian, where the electron correlation and 
the orbital degeneracy are taken into account, is introduced.  
In Sect.~III,  
formulation to calculate the phase diagram 
at finite $T$ and $x$ is presented.  
Numerical results are shown in Sect.~IV  
and the last section is devoted to summary and discussion.
\section{Model}
Let us consider the model Hamiltonian 
which describes the electronic structure 
in perovskite manganites. 
We set up the cubic lattice consisting 
of manganeses ions. 
Two $e_g$ orbitals are introduced in 
each ion and $t_{2g}$ electrons 
are treated as a localized spin $(\vec S_{t_{2g}})$ with $S=3/2$. 
Between $e_g$ electrons, 
three kinds of the Coulomb interaction,
that is, the intra-orbital Coulomb interaction ($U$), 
the inter-orbital one ($U'$) and 
the exchange interaction($I$), are taken into account. 
There also exist the Hund coupling ($J_H$) between $e_g$ and 
$t_{2g}$ spins and the electron transfer $t_{ij}^{\gamma \gamma'}$ 
between site $i$ with orbital $\gamma$ and 
site $j$ with $\gamma'$. 
Among these energies, the Coulomb interactions are the 
largest one. 
Therefore, by excluding the doubly occupied state at each site, 
we derive the effective Hamiltonian describing the 
low energy spin and orbital states: \cite{ishihara1}
\begin{equation}
{\cal H}={\cal H}_{t}+{\cal H}_{J}+{\cal H}_{H}+{\cal H}_{AF} \ . 
\label{eq:ham}
\end{equation}
The first and second terms correspond to the so-called
$t$- and $J$-terms in the $tJ$-model for $e_g$ 
electrons, respectively.  These
are given by 
\begin{eqnarray}
{\cal H}_t=\sum_{\langle i j \rangle \gamma \gamma' \sigma }
t_{ij}^{\gamma \gamma'} \widetilde d_{i \gamma \sigma}^\dagger 
\widetilde d_{j \gamma' \sigma} + H.c \ , 
\end{eqnarray}
and
\begin{eqnarray}
{\cal H}_{J}=&-&2J_1\sum_{\langle ij \rangle } 
\Bigl ( {3 \over 4} n_i n_j + \vec S_i \cdot \vec S_j   \Bigr )
\Bigl ( {1 \over 4}  - \tau_i^l \tau_j^l \Bigr ) \nonumber \\
     &-&2J_2\sum_{\langle ij \rangle } 
\Bigl ( {1 \over 4} n_i n_j  - \vec S_i \cdot \vec S_j   \Bigr )
\Bigl ( {3 \over 4}   + \tau_i^l \tau_j^l +\tau_i^l+\tau_j^l \Bigr ) \ , 
\label{eq:hj}
\end{eqnarray}
where
\begin{equation}
 \tau_i^l =\cos \Bigl({2 \pi \over 3} n_l \Bigr)
T_{iz}-\sin \Bigl({2 \pi \over 3} n_l \Bigr) T_{ix} \ , 
\end{equation}
and 
$(n_x,n_y,n_z)$$=(1,2,3)$. 
$l$ denotes the direction of bond connecting $i$ and $j$ sites.  
$\widetilde d_{i \gamma \sigma}$ is the annihilation operator of $e_g$ electron at site $
i$ with spin $\sigma$ and orbital $\gamma$ with excluding double 
occupancy. 
$\vec S_i$ is the spin operator of the $e_g$ electron  
and $\vec T_i$ is the pseudo-spin operator 
for the orbital degree of freedom defined as 
$\vec T_i=(1/2)\sum_{\sigma \gamma \gamma'} 
\widetilde d_{i \gamma \sigma}^\dagger (\vec \sigma)_{\gamma \gamma'} 
\widetilde d_{i \gamma' \sigma} $. 
$J_1=t_0^2/(U'-I)$ and $J_2=t_0^2/(U'+I+2J_H)$ 
where $t_0$ is the transfer intensity between 
$d_{3z^2-r^2}$ orbitals in the $z$-direction, and the relation 
$U=U'+I$ is assumed. 
The orbital dependence of 
$t_{ij}^{\gamma \gamma'}$ is estimated from the 
Slater-Koster formulas. 
The third and fourth terms  in Eq.~(\ref{eq:ham})
describe the Hund coupling between $e_g$ and $t_{2g}$ spins 
and the antiferromagnetic interaction between $t_{2g}$ spins, 
respectively, as expressed as 
\begin{equation}
{\cal H}_{H}=-J_H\sum_i \vec S_{t_{2g} i} \cdot \vec S_i  \ , 
\label{eq:hh}
\end{equation}
and 
\begin{equation} 
{\cal H}_{AF}=J_{AF}\sum_{\langle ij \rangle} 
\vec S_{t_{2g} i} \cdot \vec S_{t_{2g} j} \ . 
\label{eq:haf}
\end{equation}
The detailed derivation of the Hamiltonian 
is presented in Ref.\onlinecite{ishihara1}. 
Main features of the Hamiltonian are summarized 
as follows: 
1) This is applicable to the doped manganites, as well as 
the undoped insulator. 
2) Since $J_1 > J_2$, the ferromagnetic state 
associated with the AF-type orbital order 
is stabilized by ${\cal H}_J$. 
Therefore, two kinds of the ferromagnetic interaction, 
that is, SE and DE are included in the model. 
3) As seen in ${\cal H}_J$, the orbital 
pseudo-spin space is strongly 
anisotropic unlike the spin space. 
\section{Formulation}
In order to calculate the spin and orbital states at finite 
temperatures and investigate the phase separation, 
we generalize the mean field theory 
proposed by de Gennes. \cite{degennes} 
Hereafter, 
the spin ($\vec S$) and pseudo-spin ($\vec T$)
variables are denoted by $\vec u$ in the unified fashion. 
In this theory,  
the spin and orbital pseudo-spin
are treated as classical 
vectors as follows 
\begin{equation}
(S^x_i, S^y_i, S^z_i) =
{1 \over 2} (\sin \theta^s_i \cos \phi^s_i, \sin \theta^s_i \sin \phi^s_i, \cos \theta^s_i) \ , 
\end{equation}
and 
\begin{equation}
(T^x_i, T^y_i, T^z_i) = {1 \over 2} (\sin \theta^t_i, 0, \cos \theta^t_i) \ , 
\label{eq:tdef}  
\end{equation}
where the motion of the pseudo-spin is assumed to be confined in the 
$xz$-plane.
$\theta^t_i$ in Eq.~(\ref{eq:tdef}) characterizes 
the orbital state at site $i$ as  
\begin{equation}
 |\theta_i^t \rangle=\cos(\theta_i^t/2) |d_{3z^2-r^2} \rangle
                     +\sin(\theta_i^t/2) |d_{x^2-y^2} \rangle \ . 
\end{equation}
$t_{2g}$ spins are assumed to be parallel to the $e_g$ one. 
The thermal distributions of the spin and pseudo-spin 
are described by the distribution function which 
is a function of the relative angle between 
$\vec u_i$  and the mean field $\vec \lambda^u_i$,  
\begin{equation}
w^u_i (\vec u_i) = {1 \over \nu^u} \exp (\vec \lambda^u_i \cdot \vec m^u_i) \ , 
\label{eq:wst}
\end{equation}
where 
$\vec m^u (\equiv \vec u_i / |\vec u|)$ 
is termed the spin(pseudo-spin) magnetization and 
the normalization factor is defined by  
\begin{equation}
\nu^s = \int_0^\pi d \theta \int_0^{2 \pi} d \phi
\exp (\lambda^s \cos \theta) \ ,
\end{equation}
and 
\begin{equation}
\nu^t = \int_0^{2 \pi} d \theta
\exp (\lambda^t \cos \theta).
\end{equation}
The mean fields are 
assumed to be written as   
${\vec \lambda^u_i}=
\lambda^u (\sin \Theta^u_i, 0, \cos \Theta^u_i)$.
By utilizing the distribution functions defined in Eq.~(\ref{eq:wst}), 
the expectation values of operators 
$A_i (\vec S)$ and $B_i (\vec T)$ 
are obtained as 
\begin{equation}
\langle A_i \rangle_s = 
\int_0^\pi d\theta^s \int_0^{2 \pi} d \phi^s w^s_i (\vec S_i) A(\vec S) \ , 
\label{eq:avea}
\end{equation}
and 
\begin{equation}
\langle B_i \rangle_t = 
\int_0^{2 \pi} d\theta^t w^t_i (\vec T_i) B(\vec T) \ , 
\label{eq:aveb}
\end{equation}
respectively. 
In this scheme, 
the free energy is represented by summation of the expectation values of 
the Hamiltonian and the entropy of spin and pseudo-spin as follows: 
\begin{eqnarray}
{\cal F} =\langle {\cal H} \rangle - N T({\cal S}^s + {\cal S}^t) \ . 
\label{eq:free}
\end{eqnarray}
$N$ is the number of Mn ions and ${\cal S}^u$ is the entropy 
calculated by 
\begin{eqnarray}
{\cal S}^u = -\langle  \ln w^u (\vec u)\rangle_u \ . 
\end{eqnarray}
By minimizing $\cal F$ with respect to $\lambda^u_i$ and $\Theta_i$, 
the mean field solutions are obtained. 
It is briefly noticed that the above formulation gives the unphysical states 
at very low temperatures ($T<T_{neg} \sim J_{1(2)}/10$) 
where the entropy becomes negative. 
Therefore, we restrict our calculation in the region 
above $T_{neg}$.  
However, at $T=0$, 
the spin and orbital states are calculated without any trouble 
in the entropy with the assumption of the full polarizations 
of spin and pseudo-spin.
\par
Next, we concentrate on the calculation of 
$\langle {\cal H} \rangle$ 
in Eq.~(\ref{eq:free}). 
As shown in Eq.~(\ref{eq:hj}), 
${\cal H}_J$ is represented by $\vec S$ and $\vec T$. 
By introducing the rotating frame in the spin(pseudo-spin) space, 
the $z$-component of the spin (pseudo-spin) in the frame 
is given by  
\begin{equation}
\widetilde  u_i^z=
\cos \Theta^u_i  u^z_i  + \sin \Theta^u_i u^x_i \ , 
\end{equation}
which is parallel to the mean field $\vec \lambda^u$. 
Thus, $ \langle \widetilde  u_i^z \rangle_u$ is adopted 
as the order parameter which has  
the relation, 
$
\langle \widetilde u_i^z \rangle_u = {1 \over 2 } \langle \widetilde m^{u z} \rangle_u 
$. 
The spin part in ${\cal H}_J$ 
is rewritten by using 
$\langle \widetilde m^{sz} \rangle_s$ and 
the relative angle in the spin space as 
$\langle \widetilde m^{sz} \rangle_s^2 \cos(\Theta^s_i-\Theta^s_j)$. 
On the other hand, 
the orbital part includes the term $\cos(\Theta^t_i+\Theta^t_j)$,  
which originates from the anisotropy in the orbital space. 
$\cal H_{AF}$ is also rewritten by using 
$\langle \widetilde m^{sz} \rangle_s$ and $\Theta^s$ 
under the relation of 
$\langle \vec S \rangle_s = 4\langle \vec S_{t_{2g}} \rangle_s $.  
As for the transfer term ${\cal H}_t$, 
we introduce the rotating frame \cite{ishihara2}
and decompose the electron operator as 
$\widetilde d_{i \gamma \sigma}= h^\dagger_i z_{i \sigma}^s z_{i \gamma}^t$  
where $h^\dagger_i$ is a spin-less and orbital-less fermion operator 
and $z_{i \sigma}^s$ and $z_{i \gamma}^t$ 
are the elements of the unitary matrix ($U^{s(u)}$) in the spin and pseudo-spin frames, 
respectively.  
These are defined by  
\begin{equation}
U^u  = 
\pmatrix{z_{i \uparrow}^u & -z_{i \downarrow}^{u *} \cr 
z_{i \downarrow}^u & z_{i \uparrow}^{u *} } \ , 
\label{eq:unitary} 
\end{equation}
with 
$z_{i \uparrow}^s = \cos (\theta^s_i / 2) \, e^{-i\phi^s_i / 2}$ 
and 
$z_{i \downarrow}^s = \sin (\theta^s_i / 2) \, e^{i\phi^s_i / 2}$
for spin, 
and 
$z_{i \uparrow}^t = \cos (\theta^t_i / 2)$ 
and 
$z_{i \downarrow}^t = \sin (\theta^t_i / 2)$ for orbital. 
By using the form, ${\cal H}_t$ is rewritten as
\begin{equation}
{\cal H}_t =\sum_{\langle i j \rangle}
 t_{ij}^s t_{ij}^t h_i h_j^\dagger 
+ H.c. ,
\label{ht2}
\end{equation}
with 
$
t_{ij}^s= \sum_{\sigma }z_{i \sigma}^{s *} z_{j \sigma}^s 
$, 
and
$
t_{ij}^t=\sum_{\gamma \gamma'}z_{i \gamma}^{t *} t_{ij}^{\gamma \gamma'} z_{j \gamma'}^t 
$. 
%
The former gives 
$    e^{ i(\phi_i^s-\phi_j^s)/2}$
$    \cos\theta_i^s \cos\theta_j^s$
$   -e^{-i(\phi_i^s-\phi_j^s)/2}$
$    \sin\theta_i^s \sin\theta_j^s$ 
as expected from the double exchange interaction. \cite{anderson} 
By diagonalizing the energy in the momentum space, 
${\cal H}_{t}$ is given by 
\begin{equation}
{\cal H}_t
= \sum_{\vec k} \sum_{l=1}^{N_l}
\varepsilon_{\vec k}^l h_{l \vec k}^\dagger h_{l \vec k} \ , 
\label{eq:hk}
\end{equation}
where $l$ indicates the band of $h_{l \vec k}$ 
and $N_l$ is the number of the bands. 
$\varepsilon_{\vec k}^l $
corresponds to the energy of the $l$-th band. 
As a result, the expectation value of ${\cal H}_t$ per site is obtained by 
\begin{equation}
E_t  =
\Bigl \langle 
{1 \over N} \sum_{\vec k} \sum_{l=1}^{N_l}
 \varepsilon_{\vec k}^l f_F ( \varepsilon_{\vec k}^l - \varepsilon_F )  
\Bigr \rangle \ , 
\label{eq:et}
\end{equation}
which is a function of the spin and pseudo-spin angles at each site,  
$\{ \Theta^s_i \}$ and $\{ \Theta^t_i \}$, and the amplitudes of the 
mean fields, $\lambda^s$ and $\lambda^t$. 
$\varepsilon_F $ in Eq.~(\ref{eq:et})
is the fermi energy of $h_{i \vec k}$  
determined in the equation,  
\begin{eqnarray}
x = {1 \over N} \sum_{\vec k} \sum_{l=1}^{N_l}
f_F ( \varepsilon_{\vec k}^l - \varepsilon_F ) ,
\end{eqnarray}
where $f_F(\varepsilon)$ is the fermi distribution function.  
\par
%
%
%
\begin{figure}
\epsfxsize=0.7\columnwidth
\centerline{\epsffile{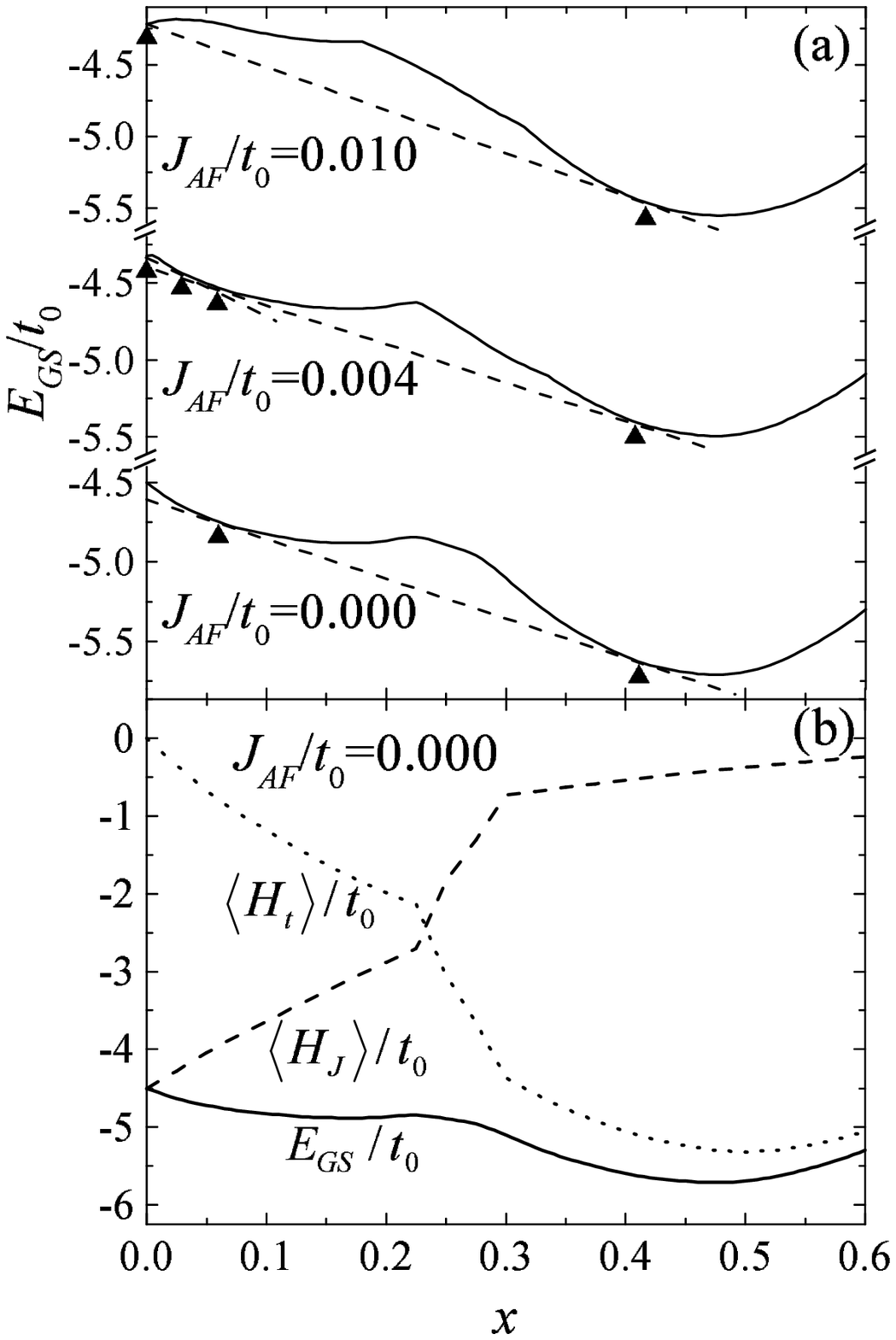}}
%
\caption{
The ground state energy $(E_{GS})$ as a function of 
hole concentration $(x)$. 
(a): $J_{AF}/t_0$ is chosen to be 0, 0.004, and 0.01.  
The broken lines and the filled triangles show  
the tangent lines of the $E_{GS}$-$x$ curve and 
the points of contact between the two. 
(b): $E_{GS}$ is decomposed into the contributions from 
$\langle {\cal H}_{t} \rangle$ and 
$\langle {\cal H}_{J} \rangle$.
$J_{AF}/t_0$ is chosen to be 0.   
The other parameter values are $J_1/t_0=0.25$, and $J_2/t_0=0.0625$. 
}
\label{fig1}
\end{figure}
%
%
\section{Numerical results}
\subsection{phase diagram at $T=0$}
%
%
\begin{figure}
\epsfxsize=0.9\columnwidth
\centerline{\epsffile{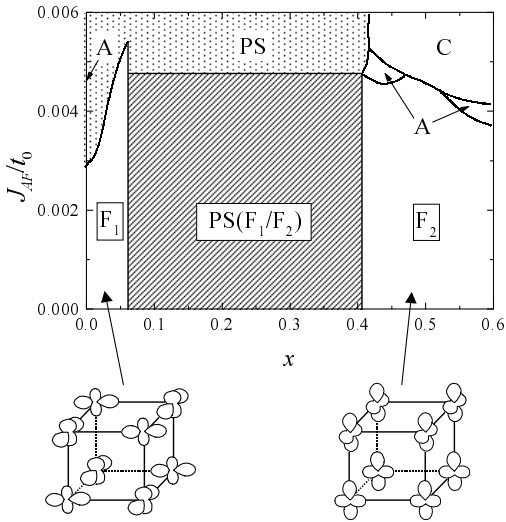}}
%
\caption{
The phase diagram at $T=0$ in the plane of antiferromagnetic 
interaction $J_{AF}$ and hole concentration $x$. 
F$_1$ and F$_2$ are the ferromagnetic phases with different 
types of orbital ordering. 
PS(F$_1$/F$_2$)  is the phase separated state between 
the F$_1$ and F$_2$ phases. 
Types of orbital ordering in the two phases are schematically 
presented. 
In the dotted region, 
there exist PS(A-AF/F$_1$) and PS(A-AF/C-AF). 
The parameter values are chosen to be $J_1/t_0=0.25$ and $J_2/t_0=0.0625$. 
}
\label{fig2}
\end{figure}
%
%
%
In this subsection, 
we show the numerical results at $T=0$. 
For examining both spin and orbital orderings, 
two kinds of sublattice are introduced. 
We assume ferromagnetic (F)-type and 
three kinds of antiferro (AF)-type 
spin (pseudo-spin) orderings, which are 
layer (A)-type, rod (C)-type and NaCl (G)-type. 
\par
In Fig.~1(a), the ground state energy $(E_{GS})$ is shown 
as a function of hole concentration ($x$) for 
several values of $J_{AF}/t_0$.  
Double- or multi-minima appear  
in the $E_{GS}$-$x$ curve depending on the value of 
$J_{AF}/t_0$. 
Therefore,  
the homogenous phase is not stable against the phase separation. 
This feature is remarkable in the region of $0.1<x<0.4$.
In Fig.~1(b), $E_{GS}$ is decomposed into $\langle {\cal H}_t \rangle$ 
and $\langle {\cal H}_J \rangle$ for $J_{AF}/t_0=0$. 
By drawing a tangent line in the $E_{GS}$-$x$ curve as shown in Fig.~1(a), 
the phase separation is obtained. 
By using the so-called Maxwell construction,  
the phase diagram at $T=0$ is obtained in the plane of 
$J_{AF}$ and $x$ (Fig.~2). 
The parameter values are chosen to be $J_1 / t_0 = 0.25$ and $J_2 / t_0 = 0.0625$.
$J_{AF}/ t_0$ for manganites 
is estimated from the N$\rm \acute{e}$el temperature in $\rm CaMnO_3$ 
to be $0.001 \sim 0.01$.
Let us consider the case of $J_{AF}/t_0 = 0.004$.
With doping of holes, 
the magnetic structure is changed as 
A-AF $\rightarrow$
PS(A-AF/F$_1$)  $\rightarrow$
F$_1$ $\rightarrow$
PS(F$_1$/F$_2$) $\rightarrow$
F$_2$, 
where PS(A/B) implies the phase separation between 
A and B phases. 
The canted spin structure does not appear. 
F$_1$ and F$_2$ are the two kinds of ferromagnetic 
phase discussed below in more detail. 
Between F$_1$ and F$_2$ phases, the PS state appears 
and dominates the large region of the phase diagram. 
For example, at $x=0.2$, the F$_1$ and 
F$_2$ phases coexist with the 
different volume fractions 
of $60\%$ and $40\%$, respectively. 
We also find the PS state between A-AF and F$_1$ phases 
in the region of $0.0<x< 0.03$. 
\par
%
%
%
\begin{figure}
\epsfxsize=0.8\columnwidth
\centerline{\epsffile{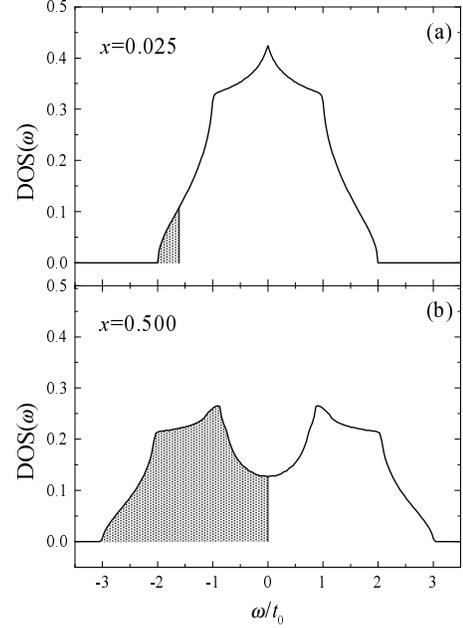}}
%
\caption{
The densities of state (DOS) for the spin-less 
and orbital-less fermions $h_{\vec k}$  
(a) in the F$_1$ phase and (b) in the F$_2$ phase. 
The shaded areas show the occupied state of $h_{\vec k}$
}
\label{fig3}
\end{figure}
%
%
Now we focus on two kinds of ferromagnetic phase 
and the PS state between them. 
The F$_1$ and F$_2$ phases originate from the SE interaction between $e_g$ orbitals 
and the DE one, respectively. 
The interactions have different types of orbital ordering as shown in 
Fig.~2. 
These are the C-type \cite{gtype} with 
$(\theta_A^t/\theta_B^t)=(\pi/2,3\pi/2)$ and 
the A-type with 
$(\theta_A^t/\theta_B^t)=(\pi/6,-\pi/6)$, 
respectively, 
where $\theta_{A(B)}^t$ is the angle in the orbital space in the $A(B)$ 
sublattice. 
It is known that 
the AF-type orbital ordering obtained in the F$_1$ phase  
is favorable to the ferromagnetic SE interaction through the 
coupling between spin and orbital degrees in ${\cal H}_J$. 
On the other hand, 
the F-type orbital ordering promotes the DE interaction by increasing the gain 
of the kinetic energy. 
To show the relation between the 
orbital ordering and the kinetic energy, 
we present the density of state (DOS) of the spin-less and orbital-less 
fermions in the 
F$_1$ and F$_2$ phases in Fig.~3(a) and (b), respectively. 
It is clearly shown that the band width in the F$_2$ phase  
is larger than that in the F$_1$ phase. 
In addition, DOS in the F$_2$ phase has a broad peak 
around $-2<\omega/t_0<-0.8$ which results 
from the quasi-one dimensional orbital ordering. 
Because of the structure in DOS, 
the kinetic energy further decreases in the F$_2$ phase more than the F$_1$ phase. 
\par
%
%
\begin{figure}
\epsfxsize=0.9\columnwidth
\centerline{\epsffile{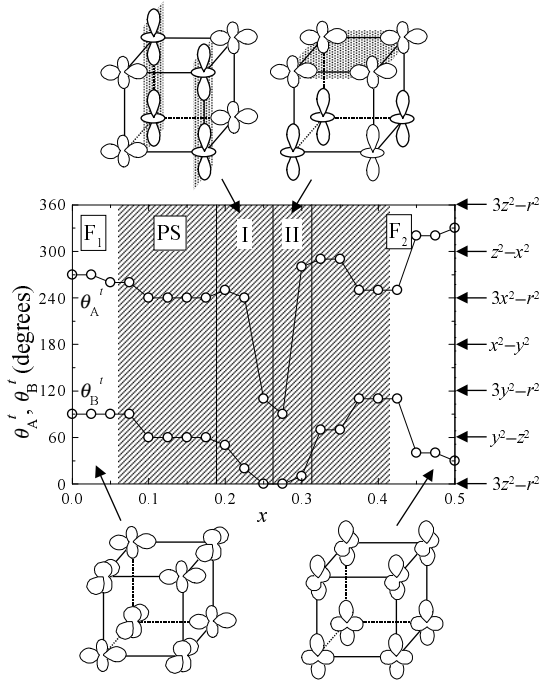}}
%
\caption{
A sequential change of orbital states as a function of hole concentration $x$. 
$\theta^t_{A(B)}$ is the angle in the orbital space in the $A(B)$ orbital sublattice. 
The schematic orbital state are shown. 
In the phase-$I$ and -$II$, 
the dotted areas show the region where the hole concentration is rich.
}
\label{fig4}
\end{figure}
%
%
In order to investigate the 
stability of the PS state appearing 
between the F$_1$ and F$_2$ phases, 
the ground state energy is decomposed into the 
contributions from the SE interaction ($ \langle {\cal H}_J \rangle$) 
and the DE one ($\langle {\cal H}_t \rangle$) (see Fig.~1(b)). 
We find that with increasing $x$, 
$\langle {\cal H}_J \rangle$ increases and 
$\langle {\cal H}_t \rangle$ decreases. 
Several kinks appear in the $\langle {\cal H}_J \rangle$-$x$ and 
$\langle {\cal H}_t \rangle$-$x$ curves, 
which imply the discontinuous 
change of the state with changing $x$.  
The PS(F$_1$/F$_2$) state shown in Fig.~2 corresponds to 
the region, where 
the two ferromagnetic interactions compete with each other 
and the discontinuous changes appear in the 
$\langle {\cal H}_{J(t)} \rangle$-$x$ curve. 
In Fig.~4, we present the $x$ dependence of the orbital state 
assuming the homogeneous phase.  
It is clearly shown that the discontinuous change of 
$\langle {\cal H}_{J(t)} \rangle$-$x$ curve  
is ascribed to that of the orbital state. 
In particular, in the phase-I and -II, 
the symmetry of the orbital is lower than that in 
the F$_1$ and F$_2$ phases and 
the stripe-type (quasi one dimensional) and sheet-type 
(two dimensional) 
charge disproportion is realized, respectively. 
These remarkable features originate from the anisotropy 
in the orbital pseudo-spin space. 
We also note that because of the anisotropy, 
the orbital state dose not change continuously from 
F$_1$ to F$_2$. 
It is summarized that the main origin of the PS state 
in the ferromagnetic state is 
1) the existence of two kinds of ferromagnetic interaction 
which favor the different types of orbital state, 
and 2) the discontinuous change of orbital state  
due to the anisotropy in the orbital space unlike 
the spin case. 
%
%
%
\begin{figure}
\epsfxsize=0.85\columnwidth
\centerline{\epsffile{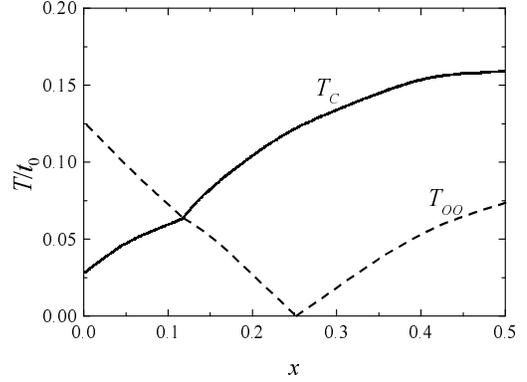}}
%
\caption{
The phase diagram in the plane of 
temperature $(T)$ and  
hole concentration $(x)$. 
The homogeneous state is assumed. 
The straight and dotted lines show the ferromagnetic 
Curie temperature $(T_C)$ and the orbital ordered 
temperature ($T_{OO}$), 
respectively. 
The parameter values are chosen to be 
$J_1/t_0=0.25$, $J_2/t_0=0.0625$ and 
$J_{AF}/t_0=0.004$. 
}
\label{fig5}
\end{figure}
%
%
\subsection{phase diagram at finite $T$}
In this subsection, 
we show the numerical results at finite $T$ 
and discuss how the PS state changes with $T$. 
As the order parameter of spin, 
we assume the ferromagnetic ordering 
and focus on the F$_1$ and F$_2$ phases and the PS state 
between them. 
We consider the G- and F-type orbital orderings 
which are enough to discuss the orbital state 
in the ferromagnetic state of the present interest. 
\par
In Fig.~5, 
the phase diagram is presented at finite $T$ 
where the homogeneous phase is assumed. 
Parameter values are chosen to be $J_{AF} / t_0 = 0.004$, 
$J_1/t_0=0.25$ and $J_2/t_0=0.0625$. 
At $x=0.0$, 
the orbital ordered temperature ($T_{OO}$) 
is higher than the ferromagnetic Curie temperature ($T_C$), 
because the interaction between 
orbitals $(3J_1/2)$ in the paramagnetic state 
is larger than that between spins $(J_1/2)$ in the orbital 
disordered state, 
as seen in the first term in ${\cal H}_{J}$.  
With increasing $x$, 
$T_C$ monotonically increases. 
On the other hand, $T_{OO}$ decreases and 
becomes its minimum around $x \sim 0.25$. 
This is the consequence of the change of orbital ordering 
from G-type 
to F-type.
The G- and F-type orbital orderings are favorable to   
the SE and DE interactions, respectively, 
so that the orderings occur in the lower and higher $x$ regions. 
In Fig.~6(a), 
we present the free energy as a function of $x$ 
at several temperatures.  
For $T/t_0 < 0.025$, 
the double minima around $x=0.1$ and $0.4$ exist 
as discussed in the 
previous subsection at $T=0$. 
With increasing $T$, 
the double minima are gradually smeared out 
and a new local minimum appears around $x=0.3$. 
It implies that another phase becomes stable around $x=0.3$  
and two different kinds of the PS state appear
at the temperature. 
With further increasing temperature, 
several shallow minima appear in the ${\cal F}$-$x$ curve. 
Finally, the fine structure disappears and 
the homogeneous phase becomes stable in the whole region of $x$. 
In Fig.~6(b), the free energy is decomposed into the 
contributions from $T{\cal S}$, 
$\langle {\cal H}_t \rangle $ and 
$\langle {\cal H}_J \rangle$ at $T/t_0=0.04$. 
\par
%
%
\begin{figure}
\epsfxsize=0.8\columnwidth
\centerline{\epsffile{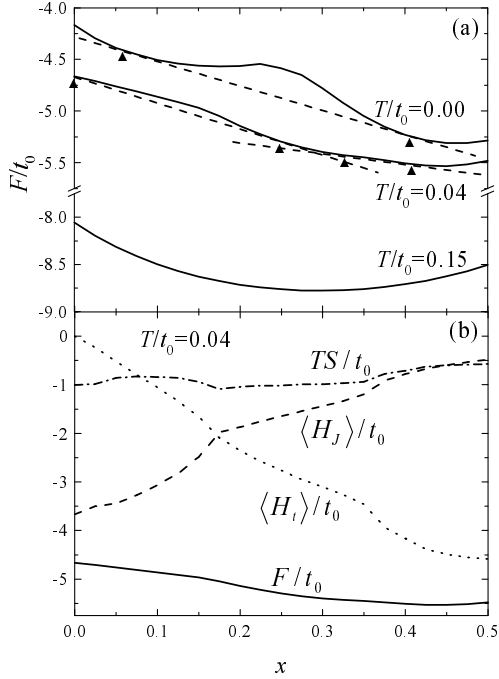}}
%
\caption{
The free energy as a function of hole concentration $(x)$. 
(a): $T/t_0$ is chosen to be 0, 0.04, and 0.15. 
The broken lines and the filled triangles show  
the tangent lines of the $\cal F$-$x$ curve and 
the points of contact between the two. 
(b):  $\cal F$ is decomposed into the contributions from 
$T{\cal S}$, 
$\langle {\cal H}_{t} \rangle$ and 
$\langle {\cal H}_{J} \rangle$.  
$T/t_0$ is chosen to be 0.04. 
}
\label{fig6}
\end{figure}
%
%

%
\begin{figure}
\epsfxsize=0.85\columnwidth
\centerline{\epsffile{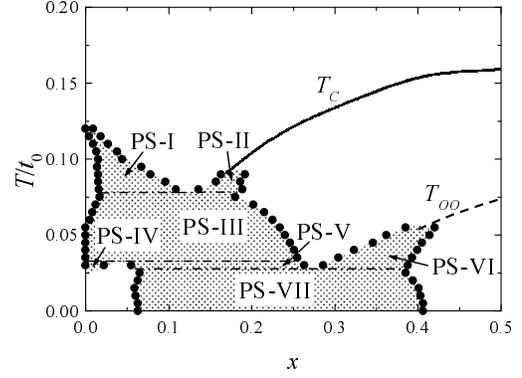}}
%
\caption{
The phase diagram 
at finite temperatures. 
The shaded area shows the phase separated region. 
The spin and orbital states in each 
state is 
PS-I:   PS(spin-P, orbital-G /spin-P, orbital-P), 
PS-II:  PS(spin-P, orbital-P /spin-F, orbital-P), 
PS-III: PS(spin-P, orbital-G /spin-F, orbital-P), 
PS-IV:  PS(spin-P, orbital-G /spin-F, orbital-G), 
PS-V:   PS(spin-F, orbital-G /spin-F, orbital-P), 
PS-VI:  PS(spin-F, orbital-P /spin-F, orbital-F), and 
PS-VII: PS(spin-F, orbital-G /spin-F, orbital-F)=(F$_1$/F$_2$). 
The parameter values are the same as those in Fig.5. 
}
\label{fig7}
\end{figure}
%
%
By applying the Maxwell construction 
to the free energy presented in Fig.~6(a), 
the PS states are obtained and presented in Fig.~7. 
The PS states dominate the large area in the $x$-$T$ plane. 
A variety of the PS states appears with 
several types of spin and orbital states. 
Each PS state is represented 
by the combination of spin and orbital states, such as   
PS(spin-P, orbital-G/spin-F, orbital-P) for PS-III  
and 
PS(spin-F, orbital-G/spin-F, orbital-F)=PS(F$_1$/F$_2$) 
for PS-VII. 
Here, P indicates the paramagnetic (orbital) state. 
It is mentioned that the phase diagram in Fig.~7  
has much analogy with that in eutectic alloys. 
For example, let us focus on the region below $T/t_0=0.05$. 
Here, the F$_1$ and F$_2$ phases and PS-VII 
correspond to the two kinds of 
homogeneous solid phases, termed A and B,  
and the PS state between them (PS(A/B)) in binary alloys, 
respectively. 
In the case of the binary alloys, 
the liquid(L)-phase becomes stable due to the entropy 
at high temperatures.  
Thus, with increasing temperature,   
the successive transition occurs as 
PS(A/B)
$\rightarrow$
(PS(L/A(B)))
$\rightarrow$
L. 
The states, 
L, PS(L/A) and PS(L/B), correspond to 
the (spin-F, orbital-P) phase, PS-V, and PS-VI in Fig.~7, 
respectively. 
By the analogy between two systems, the point at 
$T/t_0=0.025$ and $x=0.27$ corresponds to the eutectic point. 
In the ${\cal F}$-$x$ curve shown in Fig.~6, 
above three states reflect on the three minima 
observed at $T/t_0=0.004$. 
By decomposing the free energy into the three terms: 
$ \langle {\cal H}_{J} \rangle $, 
$ \langle {\cal H}_{t} \rangle $ and $T{\cal S}$, 
we confirm that the middle part corresponding to the 
(spin-F, orbital-P) phase
is stabilized by the entropy. 
\par
In Fig.~8, we present effects of the magnetic field ($B$)
on the phase diagram. 
The magnitude of the applied magnetic field is chosen to be 
$g {\mu}_{B} B / t_0 = 0.02$ which corresponds to 
$50$ Tesla for $t_0=0.3eV$ and $g=2$.
We find that the PS state shrinks in the magnetic field. 
The remarkable change is observed 
in PS-II and III where 
the spin-F and -P phases coexist. 
The magnetic field stabilizes the ferromagnetic phase 
so that the PS states are replaced by PS-V and the 
uniform ferromagnetic state. 
The region of PS-VII (PS(F$_1$/F$_2$)) 
is also suppressed in the magnetic field.  
Because the magnitude of the magnetization in the phase F$_1$ 
is smaller than that in the F$_2$ phase, 
the magnetic field increases the magnetization and 
stabilizes the F$_1$ phase.  
%
%
%
\begin{figure}
\epsfxsize=0.85\columnwidth
\centerline{\epsffile{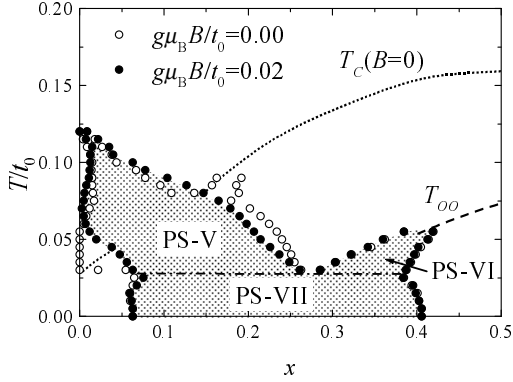}}
%
\caption{
The phase diagram at finite temperatures in 
the applied magnetic field ($B$). 
The open and filled circles show the 
boundary of the phase separated region in 
$g\mu_B B/t_0=0$ and $0.02$, respectively.  
The other parameter values are the same as those in Fig.5.
}
\label{fig8}
\end{figure}
%
%
\section{summary and discussion}
In this paper, we study the spin and orbital phase diagram for 
perovskite manganites  at finite $T$ and $x$. 
In particular, we pay our attention to 
two kinds of ferromagnetic phase appearing 
at different hole concentrations. 
The SE and DE interactions dominate the ferromagnetic phases
in the lower and higher $x$ and 
favor the AF- and F-type 
orbital orderings, respectively. 
Between the phases, 
the two interactions compete with each other and 
the phases are unstable against the phase separation. 
The PS states at finite $T$  
have much analogy with that in the binary alloys. 
\par
It is worth to compare PS(F$_1$/F$_2$)
with PS(AF/F). 
As shown in Fig.~2 ($J_{AF}/t_0=0.004$), 
PS(F$_1$/F$_2$) appears in the region of 
higher $x$ than PS(A-AF/F). 
This originates from the following 
sequential change of the state with doping of holes as  
I:(spin-A,orbital-G)
$\rightarrow$
II:(spin-F,orbital-G)
$\rightarrow$
III:(spin-F, orbital-F). 
The orbital state changes at higher $x$ 
than the spin state. 
As a result, PS(A-AF/F) and PS(F$_1$/F$_2$) appear between 
I and II, and II and III, respectively. 
This is because 1) at $x=0$, 
the ferromagnetic interaction between spins is weaker than 
the AF one between orbitals, as mentioned in Sect.~IV B, 
and 2) at $x=0$, the AF interaction along the $c$-axis is 
much weaker than the ferromagnetic one in the $ab$-plane. 
We also notice in Fig.~2 that 
PS(F$_1$/F$_2$) dominates a larger region in the phase diagram 
than PS(A-AF/F). 
This mainly results from 
the anisotropy in the orbital 
pseudo-spin space. 
As shown in Fig.~4, 
$\theta^t_{A(B)}$ indicating the orbital state discontinuously changes 
with $x$ in the region of $0.06<x<0.41$. 
Continuous change from F$_1$ to F$_2$  
is prevented by the anisotropy in the orbital space. 
This is highly in contrast to the spin case 
where the incommensurate and/or flux states associated with the 
continuous change of the spin angle become more stable than 
some PS states. \cite{inoue}
The anisotropy in the orbital space also 
stabilizes the homogeneous state  
in the region of $x<0.06$. 
On the other hand, PS(A-AF/F) appears  
by doping of infinitesimal holes. 
Furthermore,  
the microscopic charge segregation appearing in the phase-I and -II (Fig.~4)  
is also due to the orbital degree of freedom. 
Here, the stripe- or sheet-type charge disproportion 
is realized and the SE and DE interactions dominate 
different microscopic regions (bonds). 
These unique phases are ascribed to the dimensionality 
control of charge carriers through the orbital orderings. 
It is mentioned that when the orbital degree of freedom is taken into account, 
PS(AF/F) \cite{nagaev} discussed in the 
double exchange model \cite{yunoki1} is suppressed. 
This is because A-AF is realized at $x=0$ 
instead of G-AF and 
the ratio of the band width between A-AF and F 
is $W_{AF}/W_F$=2/3. 
This ratio is much larger than that between G-AF and F which 
is of the order of $O(t_0/J_H)$.  
Therefore, the PS region, where the compressibility 
$(\kappa=(\partial \mu/\partial x)^{-1})$
is negative, shrinks. 
The $(d_{3x^2-r^2}/d_{3y^2-r^2})$-type orbital ordering 
expected from the lattice distortion in $\rm LaMnO_3$
further enhances $W_{AF}/W_F$, because the transfer intensity along the $c$-axis 
is reduced in the ordering. 
\par
It should be noticed that the 
following effects may suppress the phase separation 
discussed in the paper. 
In the present calculation, 
the order parameters for spin and orbital 
are restricted in a dice consisting of $2 \times 2 \times 2$ Mn ions. 
Other types of the ordering become candidates 
for the solution with the lower energy, especially, 
in the lightly doped region. 
However, the orbital ordering with the long periodicity 
is less important in comparison 
with that in the spin case. \cite{inoue}
The orbital ordering associated with  
continuous change of the pseudo-spin is prohibited by the anisotropy, 
as discussed above. 
Neither the quantum fluctuation neglected in the mean field 
theory nor the long range Coulomb interaction 
favor the phase separation. 
When the effects are taken into account, 
the area of PS in the $x$-$T$ plane shrinks
and certain regions will be replaced by the homogeneous phases. 
In this case, 
it is expected that the phases with 
the microscopic charge segregation, such as the phase-I 
and -II shown in Fig.~4, remain, instead of the 
macroscopic phase separation.   
\par
For observation of the PS(F$_1$/F$_2$) state 
proposed in this paper, 
the most direct probe is the resonant x-ray scattering 
which has recently been developed as a technique to observe 
the orbital ordering. \cite{murakami,ishihara3}
Here, the detailed measurement at several orbital reflection points 
are required to confirm PS where different orbital orderings coexist. 
Observation of the inhomogeneous lattice distortion 
is also considered as one of the evidence of PS(F$_1$/F$_2$), 
although this is an indirect one. 
Several experimental results have  
reported an inhomogenenity in the lattice degree of freedom.  
In La$_{1-x}$Sr$_x$MnO$_3$, 
two kinds of Mn-O bond with different lengths are observed by 
the pair distribution function analyses. \cite{louca}
These values are almost independent of $x$, 
although the averaged orthohombicity decreases with $x$. 
Since two kinds of the bond are observed far below $T_C$ 
where the magnetization is almost saturated, 
PS(AF/F) is excluded and 
PS with different orbital orderings 
explain the experimental results. 
The more direct evidence of PS 
was reported by the synchrotron x-ray diffraction in 
$\rm La_{0.88}Sr_{0.12}MnO_3$. \cite{cox}
Below 350K, some of the diffraction peaks split and the 
minor phase with 20$\%$ volume fraction appears. 
This phase shows the larger orthohombic distortion 
than the major one in the region of $105{\rm K}<T<350{\rm K}$. 
Thus, the experimental data are consistent 
with the existence of PS(F$_1$/F$_2$) where the major and minor 
phases correspond to the F$_2$ and F$_1$ phases, respectively.  
In this compound, 
the first order phase transition from  
ferromagnetic insulator to ferromagnetic metal 
occurs at $T=145{\rm K}$. \cite{endoh}
Through the systematic experiments, 
it has been revealed that this magnetic transition is ascribed 
to the transition between the orbital ordering and disordering. 
The experimental results strongly suggest that 
the two different interactions, i.e., SE and DE,  
are concerned in the transition and 
unconventional experimental results are understood in terms of 
the interactions. 
It is desired to carry out further experimental and theoretical 
investigations to clarify roles of the PS state on the unconventional phenomena. 
\begin{acknowledgments}
Authors would like to thank 
Y. Endoh, K. Hirota and H. Nojiri 
for their valuable discussions.  
This work was supported by the Grant in Aid from Ministry of Education, 
Science and Culture of Japan, CREST and NEDO. 
S.O. acknowledges the financial support of JSPS Research Fellowships 
for Young Scientists. 
Part of the numerical calculation was performed in the HITACS-3800/380 
superconputing facilities in IMR, Tohoku University. 
\end{acknowledgments}


\begin{references}
\bibitem{gtype}
The G-type orbital ordering is degenerate with the C-type ordering.
\end{references}
\end{document}